\documentclass{llncs}
\usepackage{makeidx}  
\usepackage{graphicx} 
\usepackage{subeqnar} 
\usepackage{multicol} 

\usepackage{amsmath}
\usepackage{amssymb}
\usepackage{amscd}
\usepackage{latexsym}
\usepackage{epsfig}

\newtheorem{defi}{Definition}

\newtheorem{Theorem}[defi]{Theorem}

\newtheorem{Lemma}[defi]{Lemma}

\date{}

\begin{document}

\pagestyle{plain}  

\title{Exact quantum query complexity for total Boolean functions}
\titlerunning{Exact quantum query complexity for total Boolean functions}

\author{Gatis Midrij\= anis
\thanks{Research supported by Grant No.01.0354 from the
Latvian Council of Science, and  Contract IST-1999-11234 (QAIP)
from the European Commission.}}
\authorrunning{Gatis Midrij\= anis }

\institute{University of Latvia, Rai\c na bulv\= aris 19, Riga,
Latvia. Email: \texttt{gatis@zzdats.lv}. Fax:
\texttt{+371-7820153}.}

\maketitle

\begin{abstract}
We will show that if there exists a quantum query algorithm that
exactly computes some total Boolean function $f$ by making $T$
queries, then there is a classical deterministic algorithm
$\mathcal{A}$ that exactly computes $f$ making $O(T^3)$ queries.
The best know bound previously was $O(T^4)$ due to Beals et al.
~\cite{Beals}.

\end{abstract}

\section{Introduction, motivation and results}
\label{sec:Intro}

The laws of the quantum world offers to construct new models of
computation that possibly are more adequate to nature. The one of
the most popular models of quantum computing is quantum query
algorithms. In this paper we will view only quantum query
algorithms computing total Boolean functions. There are some very
exciting quantum query algorithms that are better than their
classical analogs. The one example is Grover's search
algorithm~\cite{Grover} that computes OR function with probability
$2/3$ making $O(\sqrt n)$ queries, where $n$ is number of Boolean
variables. The other example is exact (giving right answer with
probability $1$) quantum algorithm for PARITY making $n/2$ queries
\cite{DeutschJozsaXOR}. It is the best from known exact quantum
query algorithms for total Boolean functions.

Those amazing examples show that proving nontrivial lower bounds
for quantum algorithms are essentially necessary. A lot of work
has been done on it, however many problems are still open.

We will focus on exact quantum query algorithms. There are two
general methods how to show quantum lower bounds. The first is
adversary method (the survey and the most general version can be
found in paper of Laplante and Magniez~\cite{AdverKolmog}). The
second is quantum query lower bound by polynomials introduced by
Beals et al.~\cite{Beals}. Their power is incomparable, see for
example \cite{AmbPolyVScompl}. Beals et al.~\cite{Beals} showed
that the number of queries needed to compute a Boolean function
$f$ by a quantum algorithm exactly $Q_E(f)$ is at least
$deg(f)/2$, where $deg(f)$ is the degree of multilinear polynomial
representing $f$. Nisan and Smolensky~\cite{NisanSmol} showed that
the number of queries needed to compute $f$ by a deterministic
algorithm $D(f)$ is at most $2deg(f)^4$. It implies $D(f) \leq
32Q_E(f)^4$.

In this paper we will show that $D(f) \leq 2deg(f)^3$ thus
deriving $D(f) \leq 16Q_E(f)^3$.

The best known result from the opposite direction is $D(f) =
deg(f)^{log_3 6}$ by Kushilevitz~\cite{Kushil}. The other is $D(f)
= deg(f)^{log_2 3}$ by Nisan and Szegedy~\cite{NisanSzegedy} and
Ambainis~\cite{AmbPolyVScompl}.

\section{Preliminaries}
\label{sec:Preliminaries}

\subsection{Quantum query algorithms}
\label{sec:Preliminaries:query_alg}

A good survey on decision tree complexity is by Buhrman and de
Wolf~\cite{BuhrWolfSurvey}. We will give only brief summary on
definition.

We consider computing a Boolean function $f(x_1,...,x_N):\{0,1\}^N
\rightarrow \{0,1\}$ in the quantum query model. In this model,
the input bits can be accessed by queries to an oracle $X$ and the
complexity of $f$ is the number of queries needed to compute $f$.
A quantum computation with $T$ queries is just a sequence of
unitary transformations
$$U_1 \rightarrow O \rightarrow U_2 \rightarrow O \rightarrow ...
\rightarrow U_{T-1} \rightarrow O \rightarrow U_{T} \rightarrow
O.$$

$U_j$ can be arbitrary unitary transformation that do not depend
on the input bits $x_1,...,x_N$. $O$ are query transformations. To
define $O$, we represent basis states as $| i, b, z \rangle$ where
$i$ consists of $\lceil logN \rceil$ bits, $b$ is one bit and $z$
consists of all other bits. Then, $O$ maps $|i, b, z \rangle$ to
$(-1)^{b·x_i} |i, b, z \rangle$ (i.e., we change phase depending
on $x_i$). The computation starts with a state $|0 \rangle$. Then,
we apply $U_1, O, . . ., O, U_T$ and measure the final state. The
result of the computation is the rightmost bit of the state
obtained by the measurement. The quantum computation computes $f$
exactly if, for every $x = (x_1, ..., x_N)$, the rightmost bit of
$U_TO_x . . .O_xU_1|0 \rangle$ equals $f(x_1, . . . , x_N)$ with
certainty. $Q_E(f)$ denotes the minimum number $T$ of queries in a
quantum algorithm that computes $f$ exactly.

\subsection{Quantum query lower bounds}
\label{sec:Preliminaries:lower}

To see quantum and randomized query lower bounds by adversary
method one can start with \cite{AdverKolmog}. We will use
polynomials method, derived by Nisan and
Szegedy~\cite{NisanSzegedy} and Beals et al.~\cite{Beals}. Quite
often it is used to derive quantum lower bounds, for example in
~\cite{AaronsonCertific}, ~\cite{AaronsonColl}, ~\cite{AmbCol},
~\cite{BuhrWolfZeroComposed}, ~\cite{BuhrZeroBounds},
~\cite{Kutin}, ~\cite{MidrijanisSetEqual}, ~\cite{ShiAprox},
~\cite{ShiColl}, ~\cite{ShiLinear}, ~\cite{WolfND}.

For any Boolean function $f$, there is a unique multilinear
polynomial $g$ such that $f(x_1,..., x_N)=g(x_1,...,x_N)$ for all
$x_1,...,x_N \in \{0,1\}$. We say that $g$ $represents$ $f$. Let
$deg(f)$ denote the degree of $g$. It is known that

\begin{Theorem}
\label{thm:QuantZerErr} \cite{Beals} For any total Boolean $f$,
$Q_E(f) \geq deg(f)/2$.
\end{Theorem}

The block sensitivity of $f$ on $x$ is the maximum number of
disjoint $B_j\subseteq \{1, \ldots, n\}$ such that $f(x^{B_j})\neq
f(x)$, $x^{B_j}$ being $x$ with all $x_i$ for $i\in B_j$ changed
to $1-x_i$. We denote it $bs_x(f)$. Let $bs(f)=\max bs_x(f)$. It
is known that

\begin{Theorem}
\label{thm:BlockSensVSDegree} \cite{NisanSzegedy} For any total
Boolean function $f$, $bs(f) \leq 2deg(f)^2$.
\end{Theorem}

\section{Deterministic vs. quantum exact algorithms}\label{exact}

Now we will show that $D(f)$ is upper bounded by $2deg(f)^3$ for
every Boolean function $f$. Our method will be quite similar to
Nisan and Smolensky \cite{NisanSmol}. Sometimes we will think
about Boolean function $f$ as polynomial representing it. Here
$maxonomial$ of polynomial $f$ is a monomial with maximal degree.

\begin{Lemma}\label{lemma:maxonBS}
    For every word $w \in \{0,1\}^N$ and every maxonomial $M$ of
    $f$, there is a set $B$ of variables in $M$ such that $f(w^B) \neq
    f(w)$.
\end{Lemma}
\proof Obtain restricted function $g$ from $f$ by setting all
variables outside of $M$ according to $w$. This $g$ contains
monomial $M$ therefore it cannot be constant. Obtain word $w' \in
\{0,1\}^{|M|}$ that assigns values from $w$ to variables in $M$.
Thus there is some set $B$ of variables in $M$ that makes $g(w'^B)
\neq g(w')$ and hence $f(w^B) \neq f(w)$.

\qed

\begin{Theorem}\label{thm:detVSdegree}
    For every total Boolean function f holds $D(f) \leq 2 deg(f)^3$.
\end{Theorem}
\proof The deterministic algorithm $\mathcal{A}$ is written in
pseudo code, as function from polynomial $f$ and word $X \in
\{0,1\}^N$ that returns value of $f(X)$. The algorithm
$\mathcal{A}$:

\begin{minipage}[h]{12 cm}
\begin{description}
    \item[] $\{0,1\}$ function Value$\diamond$f(By value f as polynomial, by queries $X \in \{0,1\}^N$)\{
        \begin{description}
            \item[1.] $p := f$;
            \item[2.] While $p$ is not constant\{
            \begin{description}
                \item[3.] Pick maxonomial $M$ in polynomial $p$;
                \item[4.] Query X-values of $M$'s variables;
                \item[5.] Replace all queried variables in $p$ with appropriate
                constants;
            \end{description}
            \item[] \};
            \item[6.] Return p;
        \end{description}
    \item[] \};
 \end{description}
\end{minipage}

The nondeterministic "pick maxonomial" can easily be made
deterministic by choosing the the first maxonomial in some fixed
order.

It is easy to see that the algorithm $\mathcal{A}$ always returns
the right result, since polynomial $p$ always describes polynomial
$f$ on word $X$.

We will show that the cycle executes at most $bs_X(f) \leq bs(f)$
times. Let $a$ denote the number of cycle executions.

We will show that $bs_X(f) \geq a$ by induction. Bases: before
cycle is executed, $X$ has no blocks since there are no variables
queried yet. Inductive assumption: after $a-1$ executions of cycle
$X$ has at least $a-1$ disjoint blocks that take their variables
only from yet queried variables and to which $f$ is sensitive on
$X$. We will show that in the next cycle execution there exists a
block $B$ that takes its variables only from variables queried in
this cycle (therefore is disjoint with previous ones) and to which
$f$ is sensitive on $X$.

Let $M$ denote maxonomial chosen in this cycle. Let $w$ denote the
word whom holds $f(X) = p(w)$. Such exists, since $p$ is just
polynomial $f$ where some variables are replaced with constants
according to $X$. It is easy to see that for any set of variables
$B$ holds $p(w^B) = f(X^B)$. Now Lemma~\ref{lemma:maxonBS} says
that there is a set $B$ of variables in $M$ such that $p(w^B) \neq
p(w)$. Since $f(X) = p(w)$ and $f(X^B) = p(w^B)$ it follows that
$f(X) \neq f(X^B)$.

$bs_X(f) \geq a$ implies that $a \leq bs(f)$, thus the cycle is
executed at most $bs_X(f) \leq bs(f)$ times.

It is easy to see that for every maxonomial $M$ holds $|M| =
deg(p)$ and at every moment $deg(p) \leq deg(f)$, thus in every
cycle $\mathcal{A}$ makes at most $deg(f)$ queries, hence $D(f)
\leq deg(f)*bs(f)$. Theorem~\ref{thm:BlockSensVSDegree} gives
$D(f) \leq 2 deg(f)^3$. \qed

\begin{Theorem}\label{thm:quant_e}
    For every total Boolean function f holds $D(f) \leq 16Q_E(f)^3$.
\end{Theorem}
\proof By Theorem~\ref{thm:detVSdegree} and
Theorem~\ref{thm:QuantZerErr}. \qed

As noticed by Ronald de Wolf, our proof works also for
"nondeterministic polynomials", giving $D(f) \leq bs(f)*ndeg(f)$,
where nondeterministic polynomial is polynomial that takes nonzero
value whenever function is $1$. This relation also improves some
results of paper by de Wolf~\cite{WolfND}. For example, it follows
that $D(f)= O(Q_2(f)^2 NQ(f))$, where $Q_2(f)$ is bounded-error
and $NQ(f)$ nondeterministic quantum query complexity for function
$f$. See \cite{WolfND} for more precise definitions.

The same proof also works to prove average-case upper bound:
$\texttt{average}D(f) \leq \texttt{average}bs(f)*ndeg(f)$, since
the run of algorithm $\mathcal{A}$ on input word $X$ is bounded by
$bs_X(f)$ cycles.

\section{Open problems}
\begin{enumerate}
    \item It is well known that block sensitivity is not tight measure
    of exact quantum query complexity
    of all Boolean functions, see Ambainis \cite{Ambainis},
    \cite{AmbPolyVScompl}. Can one somehow use it to derive better
    quantum lower bound for any total Boolean function?
    \item Can similar arguments be used to show upper bound over
    degree of approximating function, for example $D(f) =
    O(\widetilde{deg}(f)^5)$? This question is related with quantum
    bounded-error query complexity.
\end{enumerate}

\section{Acknowledgments}
I want to thank Andris Ambainis and Ronald de Wolf about useful
comments on this paper.

\end{document}